\documentclass[prl,twocolumn]{revtex4}
\usepackage{amsmath,epsfig,bm}
\usepackage{hyperref}

\newcommand\ra{\rangle}
\newcommand\prlsec[1]{\emph{#1}\ ---}

\begin{document}

\author{M.~A.~Stephanov and Y.~Yin}
\affiliation{Department of Physics, University of Illinois, Chicago, 
Illinois 60607, USA}

\title{Chiral Kinetic Theory}

\begin{abstract}
  We derive the non-equilibrium kinetic equation describing the
  motion of chiral massless particles in the regime where it can be
  considered classically. We show that the Berry
  monopole which appears at the origin of the momentum space due to
  level crossing is responsible for the chiral magnetic and vortical
  effects.
\end{abstract}

\maketitle

\prlsec{Introduction}
The generation of non-dissipative currents in a chiral (parity
violating) system in response to an external magnetic field has
attracted significant amount of interest recently. Such an effect,
noted earlier in different contexts in
Refs.\cite{PhysRevD.22.3080,Nielsen1983389}, has been recently
proposed as an intriguing explanation for the charge-dependent
correlations in heavy-ion collisions in
Ref.\cite{Fukushima:2008xe,Fukushima:2010vw} and termed chiral
magnetic effect (CME). It has been also shown recently in
Ref.\cite{Son:2009tf} that hydrodynamics of chiral systems with
anomaly requires the presence of such currents, as well as currents
induced by the vorticity of flow -- the chiral vortical effect (CVE),
discovered earlier in a microscopic calculation in astrophysical
context in Ref.\cite{PhysRevD.20.1807} and rediscovered recently in
gauge-gravity duality calculations in
Refs.\cite{Erdmenger:2008rm,Banerjee:2008th}.

The interesting applications of these chiral transport effects involve
highly non-equilibrium conditions, such as those arising in the early
stages of the heavy-ion collisions, when the magnetic fields created
by passing ions are still strong. However, derivations of these
effects have been mostly done assuming thermal and chemical
equilibrium. The aim of this paper is to address this shortcoming of
theory.

A natural framework to study non-equilibrium conditions is a kinetic
theory. As any useful theory, it has limitations, such as assumption
of the classical motion between collisions and the weakness of the
coupling. Nevertheless, a kinetic description would undoubtedly be an
important step for our understanding of the chiral transport phenomena.

Most of the ingredients of the approach presented here can be found in
the literature on the physics of geometric phases introduced by Berry
in Ref.\cite{Berry:1984jv}.  The relevant
classical equations of motion were introduced in
Ref.\cite{PhysRevB.59.14915} (see also \cite{RevModPhys.82.1959} for a
review). The kinetic equation in the presence of the Berry curvature
have been studied, e.g., in
Refs.\cite{Wong:2011nt,PhysRevB.77.035110}. The most recent and
closely relevant applications include
Refs.~\cite{Son:2012wh,Son:2012bg}. Unrelated to the above so far, but
very important step towards kinetic description of the CME and CVE, was
made recently in Ref.\cite{Loganayagam:2012pz}.

Putting these ingredients together we 
derive the desired non-equilibrium expressions for the CME and
CVE which are, to the extent of our knowledge, new.  The key
observation of the present paper is that for Weyl fermions the Berry
curvature, being the field of a monopole, leads directly to the chiral
magnetic effect. We also point out that the chiral vortical effect can
be similarly understood by simply replacing the Lorentz force due to
the magnetic field by the Coriolis force. In the following we shall
present a reasonably self-contained derivation of these results using
a formalism somewhat complementary to traditional approaches. This
will allow us also to make connections to other field-theoretical
concepts (such as abelian projection) more familiar in the particle
theory context.

\prlsec{Kinetic equation}
\label{sec:kinetic-equation}
Kinetic equation describes the motion of particles in the
  regime where collisions are rare enough that motion between
  collisions is classical.
 In terms of the distribution function $f(t,\bm x,\bm p)$ the equation reads
\begin{equation}\label{eq:dfdtC}
  \frac{df}{dt} \equiv \frac {\partial f}{\partial t} + \frac{\partial f}{\partial \bm x} \bm {\dot x }
+ \frac{\partial f}{\partial \bm p} \bm {\dot p }= C[f].
\end{equation}
We think of a ``cloud'' of particles each of which follows its
classical trajectory $\bm x(t)$, $\bm p(t)$. As a result the
distribution evolves with time in such a way that if one follows a
local volume occupied by a set of particles along the trajectory, the
number of particles in it can only be changed by collisions.

Since collisions are not the focus of this paper we shall ignore
collision integral $C[f]$ in Eq.~(\ref{eq:dfdtC}) here.

The CME is known to be closely related to the chiral
anomaly~\cite{Nielsen1983389,Fukushima:2008xe,Son:2009tf}. On the other hand, it
is clear from the description above that the number of particles in
the phase space {\em cannot
  change}. 
How could a kinetic equation account for anomalous particle number
non-conservation?  In other words, how can {\em classical} equation
account for {\em quantum} anomaly? As we shall see below, the answer
is, in two words: Berry monopole.

\prlsec{Path integral and U(2) gauge invariance}
\label{sec:path-integral-berry}
Consider the Hamiltonian for a Weyl particle:
\begin{equation}
  \label{eq:1}
  H = \bm{ \sigma\cdot p}\,.
\end{equation}
For each momentum $\bm p$ it represents a two-state system with energy gap $2|\bm p|$.

It is more straightforward to obtain classical limit in the path
integral formulation rather than in the canonical formulation of
quantum mechanics usually employed to describe Berry connection.
Consider the transition amplitude between the two spin states $i$ and
$f$.  By inserting the sums over complete sets of eigenstates of
coordinates and spin, $|\bm x,s\ra$, and momenta and spin, 
$|\bm p,\lambda\ra$, the amplitude can
be written as a path integral
\begin{multline}
  \label{eq:fi}
\langle f| e^{iH (t_f-t_i)}|i\rangle 
\\ = 
 \left[ \int {\cal D}x{\cal D}p\, {\cal P} \exp\left\{i\int_{t_i}^{t_f}(\bm {p\cdot \dot
     x}-\bm{\sigma\cdot p})dt\right\}\right]_{fi}\,,
\end{multline}
where we need to take a matrix element $[\ldots]_{fi}$ of the path-ordered product of the
matrices $\exp\{-i\bm{\sigma\cdot p}\Delta t\}$ over each path $\bm
x(t)$, $\bm p(t)$ in the
phase space. These matrices can be thought of 
as describing the rotation of the state of the
particle in the spin space as it moves along.

The massless particles we describe have only one helicity state (the
opposite helicity state corresponds to an antiparticle).
In order to consider classical motion of such a particle we need to
diagonalize the matrix in the helicity basis and then apply the usual
method of stationary phase to determine the classical trajectory.
This diagonalization can be done at each point on the
trajectory using unitary matrix $V_{\bm p}$ such that 
\begin{equation}
V^\dag_{\bm
  p}\bm{\sigma\cdot p}V_{\bm p}=|\bm p|\sigma_3\label{eq:2}\,.
\end{equation}
If the values of momenta at two neighboring points $t_1$ and $t_2$
are $\bm p_1$ and $\bm p_2$, we insert identity matrices between the
exponential factors in the following way:
\begin{multline}
  \label{eq:VeVVeV}
\ldots  V_{\bm p_2}V_{\bm p_2}^\dag\exp\{-i\bm{\sigma\cdot p_2}\Delta t\}
V_{\bm p_2}V_{\bm p_2}^\dag 
\\\times
V_{\bm p_1}V_{\bm p_1}^\dag
\exp\{-i\bm{\sigma\cdot
    p_1}\Delta t\} V_{\bm p_1}V_{\bm p_1}^\dag\ldots 
\\
=\ldots V_{\bm p_2}\exp\{-i|\bm p_2|\sigma_3\Delta t\}
V_{\bm p_2}^\dag V_{\bm p_1}
\\\times
\exp\{-i|\bm p_1|\sigma_3\Delta t\}V_{\bm p_1}^\dag\ldots \,.
\end{multline}
If the $\Delta\bm p\equiv \bm p_2 - \bm p_1$ is small, we can
write the extra unitary rotation between the two neighboring exponents as
\begin{equation}
  \label{eq:4}
  V_{\bm p_2}^\dag V_{\bm p_1}
\approx \exp(-i \bm {\hat a}_{\bm p}\bm\cdot\Delta \bm p)\,,
\mbox{ where }
\bm{\hat a}_{\bm p}= iV_{\bm
  p}^\dag\bm\nabla_{\bm p}V_{\bm p}
\end{equation}
is a hermitian $2\times 2$ matrix.

Performing the above diagonalization along the whole trajectory and
assembling the exponents into the path integral one obtains
alternative expression for the amplitude in Eq.~(\ref{eq:fi})
\begin{multline}
  \label{eq:fi-ahat}
  \langle f| e^{iH (t_f-t_i)}|i\rangle 
= 
 \left[ V_{\bm p_f}\int {\cal D}x{\cal D}p
\right.\\\left. \times {\cal P} 
\exp\left\{i\int_{t_i}^{t_f}(\bm {p\cdot \dot x}
-|\bm p|\sigma_3 -\bm{\hat a}_{\bm p}\bm\cdot \bm{\dot p})dt 
\right\} V_{\bm p_i}^\dag \right]_{fi}\,.
\end{multline}
 If we did not insist on diagonalizing the matrix $\bm{ \sigma\cdot p}$, we could
have chosen an arbitrary $U(2)$ rotation, say $V_{\bm p} U_{\bm
  p}$, instead of $V_{\bm p}$.
This results in a local ``gauge transformation'' of the ``action'' such that
\begin{equation}\label{eq:U2-gauge}
  -|\bm p|\sigma_3\to -|\bm p|U_{\bm p}^\dag\sigma_3 U_{\bm p},
\quad
\bm{\hat a}_{\bm p}\to U_{\bm p}^\dag\bm{\hat a}_{\bm p}U_{\bm p}+iU_{\bm
  p}^\dag\bm\nabla_{\bm p}U_{\bm p}\,.
\end{equation}
This gauge freedom corresponds to the free choice of the phase and
spin quantization direction for the momentum states: $|\bm p, s\rangle
\to U_{\bm p}|\bm p, s\rangle$ along the trajectory. Clearly this
choice only affects the expression for the amplitude in
Eq.~(\ref{eq:fi-ahat}), and not the value of the amplitude itself. We
use this redundancy of description to choose the helicity basis at each
$\bm p$. This choice diagonalizes $\bm{\sigma\cdot
  p}$ and enables us to take the classical limit.

\prlsec{Abelian projection and Berry monopole}
\label{sec:abel-proj-berry}
Fixing this nonabelian U(2) gauge freedom by diagonalizing the
Hamiltonian is mathematically similar to the abelian projection
introduced by 't Hooft in Ref.\cite{Hooft1981455}. In the
classical regime the contribution of the transitions
caused by the
off-diagonal components of $\bm {\hat a}_{\bm p}$ is negligible (in
Ref.\cite{Hooft1981455} the ``non-abelian'' part of the gauge field is
non-propagating due to confinement).
 Suppressing these off-diagonal
components, we still have a U(1)$\times$ U(1) gauge freedom
corresponding to selecting arbitrarily the complex phases for the
helicity eigenstates at each momentum. Focusing on helicity +1 we can
denote the corresponding diagonal component $[\bm{\hat a}_{\bm
  p}]_{11}\equiv \bm a_{\bm p}$. Then the classical action for the
helicity $+1$ particle becomes
\begin{equation}
  \label{eq:I}
 I = \int_{t_i}^{t_f}(\bm {p\cdot \dot
     x}-|\bm p| -\bm{ a}_{\bm p}\bm\cdot \bm{\dot p})dt\,.
\end{equation}
The classical, or adiabatic, approximation will break down when the
two eigenvalues of the Hamiltonian are degenerate, i.e., at $\bm p
=0$. As we shall see, this point is the source of the
effects of the quantum anomaly.

As in the 't Hooft's original application of the abelian projection,
even if the non-abelian field $\bm{\hat a}_{\bm p}$ is a pure gauge,
Eq.~(\ref{eq:4}), the abelian component $[\bm{\hat a}_{\bm
  p}]_{11}\equiv \bm a_{\bm p}$ is non-trivial. Finding the unitary
matrix $V_{\bm p}$ in Eq.~(\ref{eq:2}) and calculating $\bm{\hat
  a}_{\bm p}$ in Eq.~(\ref{eq:4}), one obtains the well-known result
that the corresponding abelian field $\bm a_{\bm p}$ is the field of a
``monopole''~\cite{Hooft1974276,Polyakov:1974ek} at $|\bm p|=0$.  Of
course, the physical amplitude cannot depend on the gauge choice in
Eq.~(\ref{eq:I}). We expect physical observables to depend only on the
abelian field strength $\bm b = \bm\nabla_{\bm p}\bm {\times a}_{\bm
  p}$. One finds
\begin{equation}
  \label{eq:b}
  \bm b = \frac {\bm {\hat p}}{2|\bm p|^2},
\qquad\mbox{where}\quad \bm{\hat p} \equiv \frac{\bm p}{|\bm p|}\,.
\end{equation}

\prlsec{Equations of motion}
Before we write the classical equations of motion, let us quantify the
conditions of their applicability. The classical, or adiabatic, approximation
requires the off-diagonal components of $\bm{\hat a_p\cdot \dot p}$ in
Eq.~(\ref{eq:fi-ahat}) to
be small compared to the energy gap $2|\bm p|$. This means the forces,
$\bm{\dot p}$, on the particle cannot be too strong. For example, if
the particle moves in a magnetic field $B$: $B\ll |\bm
p|^2$, where we used $|\bm{\hat a_p}|\sim 1/|\bm p|$
(cf. Eq.~(\ref{eq:b})). This condition is obvious physically, since
particles with momenta as low as the momenta on the lowest Landau
orbit cannot be treated classically.

It is easy to couple the classical particle described by the action in
Eq.~(\ref{eq:I}) to external electromagnetic field given by
scalar and vector
potentials $\Phi$ and  $\bm A$. By variations of the resulting action
\begin{equation}
  \label{eq:IA}
 I = \int_{t_i}^{t_f}(\bm {p\cdot \dot 
     x}+ \bm{ A \cdot \dot x} -\Phi -|\bm p| -\bm{ a}_{\bm p}\bm\cdot \bm{\dot p})dt
\end{equation}
one obtains the desired equations of motion (cf. \cite{PhysRevB.59.14915,RevModPhys.82.1959}):
\begin{align}\label{eq:xdot}
  &\bm {\dot x } = \bm{\hat p} + \bm {\dot p\times b} ;
\\\label{eq:pdot}
& \bm {\dot p } = \bm E + \bm {\dot x\times B}.
\end{align}

Without the Berry flux $\bm b$, these equations are familiar equations
for the velocity of a massless particle and the Lorentz force. Without
electromagnetic field the Berry curvature $\bm b$ drops out of the
equations, because $\bm {\dot p}=0$.

Substituting
Eq.~(\ref{eq:pdot}) into Eq.~(\ref{eq:xdot}) and solving for
$\bm {\dot x}$ one finds:
\begin{align}\label{eq:Gdotx}
&  \sqrt G\, \bm {\dot x} 
= \bm{\hat p} + \bm{E\times b}+ \bm B(\bm{\hat p\cdot b})\,;
\\ \label{eq:Gdotp}
&  \sqrt G \bm {\dot p} 
= \bm E + \bm {\hat p\times B } + \bm b  (\bm{E\cdot B})\,. 
\end{align}
Here  $G=(1+\bm {b\cdot B})^2$ is the
determinant of the $6\times 6$ matrix of coefficients in
Eqs.~(\ref{eq:xdot}),~(\ref{eq:pdot}) for $\bm {\dot x }$ and $\bm {
  \dot p}$.
Substituting into Eq.~(\ref{eq:dfdtC})
we can then obtain the desired kinetic equation for the distribution
function of such particles in the phase space.

\prlsec{Chiral magnetic effect}
\label{sec:chir-magn-effect}
It is important to take account of the fact that the invariant measure of
the phase-space integration is given by $\sqrt G \,{d^3\bm x\,
  d^3\bm p}/{(2\pi)^3}$, see e.g., Ref.\cite{Duval:2005vn}. In particular,
one can check using equations of motion~(\ref{eq:Gdotx}),~(\ref{eq:Gdotp}) and
Maxwell equations $\bm{\nabla\cdot B}=0$, $\bm{\nabla\times E}=\partial\bm B/\partial t$ that this measure obeys Liouville equation:
\begin{equation}
  \label{eq:dGdt}
  \frac{\partial}{\partial t}\sqrt G 
+ \frac{\partial}{\partial\bm x}(\sqrt G\bm{\dot x})
+  \frac{\partial}{\partial\bm p}(\sqrt G\bm{\dot p})=2\pi \bm{E\cdot
B}\,\delta^3(\bm p)\,,
\end{equation}
where the last term is due to the Berry monopole $\bm{\nabla_p\cdot
  b}=2\pi\delta^3(\bm p)$, Eq.~(\ref{eq:b}). The last term is the
effect of the
quantum anomaly which ``injects'' particle number violation into our
otherwise classical description. It is notable that this term is localized at $\bm p
=0$, where the classical description must break down due to level crossing.

The current density is given by
\begin{equation}
  \label{eq:current}
  \bm j = \int_{\bm p} \sqrt G f \bm {\dot x }\,,
\qquad\mbox{where}\quad \int_{\bm p} \equiv \int \frac{d^3\bm p}{(2\pi)^3}\,.
\end{equation}
Substituting Eq.~(\ref{eq:Gdotx}) into Eq.~(\ref{eq:current}) we find
\begin{equation}
  \label{eq:j-pEB}
  \bm j = \int_{\bm p} \sqrt G f \bm {\dot x } = 
\int_{\bm p}  f \bm{\hat p}
+
\bm {E\times} \int_{\bm p} f \bm b
+ 
\bm B \int_{\bm p}  f (\bm{\hat p\cdot b})\,.
\end{equation}
The first term gives the usual current, while the second is the
anomalous Hall current. Both vanish in a state
with isotropic momentum distribution, such as equilibrium state.
 The last term is the
desired non-equilibrium expression of the CME.

Using notations $E=|\bm p|$ and an overbar to denote average over the
unit sphere of directions of vector $\bm{\hat p}$ we can write
\begin{equation}\label{eq:jcme}
  \bm j_\textrm{CME} 
= \bm B \int_{\bm p}  f\, (\bm{\hat p\cdot b})
=  \frac{\bm B}{4\pi^2}\int_0^\infty \overline{f(E,\bm {\hat  p})} dE\,.
\end{equation}
This equation agrees with the result conjectured in
Ref.~\cite{Loganayagam:2012pz} for an isotropic distribution. In the
case of the Fermi-Dirac distribution it reproduces the well-known
results (such as $ \bm j_\textrm{CME} = \mu\bm B/(2\pi)^2$ at zero
temperature).

\prlsec{Chiral anomaly}
To find the effect of the electromagnetic anomaly we calculate the
4-divergence of the particle number current in Eq.~(\ref{eq:current}).
It is illuminating to begin the discussion by introducing the 6+1
phase space current $(\rho,\rho\bm{\dot x},\rho\bm{\dot p})$, where
$\rho=\sqrt G f$ obeys continuity equation with a source
\begin{equation}\label{eq:drho/dt}
\frac {\partial \rho}{\partial t} + \frac{\partial  (\rho \bm {\dot x })}{\partial \bm x}
+ \frac{\partial (\rho\bm {\dot p }) }{\partial \bm p}
=
2\pi \bm{E\cdot
B}\,f\,\delta^3(\bm p)\,,
\end{equation}
which follows from Eq.~(\ref{eq:dfdtC}) (with $C[f]=0$ for simplicity)
and Eq.~(\ref{eq:dGdt}).  Integrating over momentum $\bm p$ we obtain
\begin{equation}
   \label{eq:dn/dt}
   \frac {\partial n}{\partial t} + \bm{\nabla\cdot j} 
 = \frac{1}{4\pi^2}\bm{E\cdot B}\, f_{\bm 0}\,,
 \end{equation}
 where, as in Eq.~(\ref{eq:current}), $(n,\bm j)=\int_{\bm p} (\rho,\rho \bm{\dot x})$ is the 3+1
 space-time current density and $f_{\bm 0}$ is the value of the
 distribution function $f$ at $\bm p = 0$. For the Fermi-Dirac
 distribution at zero temperature and non-zero chemical potential
 $f_0=1$ and we reproduce the standard expression of the
 electromagnetic anomaly.

  Strictly speaking the above calculation is not completely legitimate
  because we integrated over the whole momentum space, including the
  singular point $\bm p = 0$ where the classical description is not
  applicable. The way to think about this equation is to exclude the
  region $|\bm p|<\Delta$ around the singularity. The value of
  $\Delta$ must be large enough so that the classical description
  applies outside it ($\Delta \gg\sqrt B$).

  Then, in the classical region $|\bm p|>\Delta$, the 6+1 phase space
  current $(\rho,\rho\bm{\dot x},\rho\bm{\dot p})$ obeys continuity
  equation. 
  I.e., the particles, in the absence of collisions, cannot be created
  or destroyed in the classical region. They can only enter or exit
  the region through the boundary of the region at $|\bm
  p|=\Delta$. Integrating the continuity equation over the classical
  region $|\bm p|>\Delta$ and defining the 3+1 current density in the
  classical region only
$(n_\Delta,\bm j_\Delta) = \int_{|\bm p|>\Delta}(\rho,\rho\bm {\dot x
})$ we find that the non-conservation of the 3+1 space-time current is matched by
the momentum-space flux into the classical region through the boundary at $|\bm p|=\Delta$:
\begin{equation}\label{eq:S-Delta}
\frac {\partial n_\Delta}{\partial t} + \bm{\nabla\cdot j}_\Delta
= \int \frac{d\bm S_{\Delta}}{(2\pi)^3} \bm{\cdot J}_{\bm p}\,,
\end{equation}
where the flux density is given by
\begin{equation}
\bm J_{\bm p}
\equiv  {\rho\bm { \dot p }}
= (\bm E + \bm{\hat p\times B})\,{f}
+ 2\pi\,{\bm{E\cdot B}}\,
f\,\frac{\bm{\hat p}}{4\pi|\bm p|^2}
\,.
\end{equation}
The first term on the right-hand side is due to acceleration
of the particles on the boundary $|\bm p|=\Delta$ which moves them in
or out of the classical region. This term gives a
negligible contribution to the total flux in Eq.~(\ref{eq:S-Delta}) if
$\Delta$ is small enough that the variation of $f$ over the boundary
can be neglected.  The total flux from the last term, however, tends to a
finite limit when $\Delta\to 0$, which is given by Eq.~(\ref{eq:dn/dt}). The
origin of this net flux is the anomaly which operates, as is well-known,
at the point of level crossing $\bm p=0$, lying inside the region
$|\bm p|<\Delta$, where the motion of particles must be treated fully
quantum-mechanically.

\prlsec{Chiral vortical effect}
\label{sec:chir-vort-effect}
To describe chiral vortical effect we need to realize that, unlike the
external magnetic field $\bm B$, which we can set directly, the
vorticity $\bm\omega$ is a property of the flow of particles, which is
indirectly controlled by external fields and initial
conditions. Moreover, the definition of vorticity involves
hydrodynamic limit, which puts additional conditions on
flow. However, we can generalize the vorticity to non-equilibrium
flows in the following way. We can decide to observe a given local
fluid element in a co-moving frame, which will have to rotate with
angular velocity $\bm\omega$ with respect to the laboratory. The
particles will experience additional non-inertial forces in this
frame. At this point we can generalize the problem to non-equilibrium
by asking what additional currents such non-inertial forces induce.

To linear order the only such force is the Coriolis force:
\begin{equation}
  \label{eq:Corioulis}
     \bm {\dot p } = 2 |\bm p| \bm {\omega \times \dot x} + {\cal
       O}(\omega^2)
\,.
\end{equation}
(This classical result can be also verified by considering Weyl
Hamiltonian in the rotating frame.) The effect of this force is the
same as of a ``magnetic field'' $\bm B\to 2|\bm p|\bm\omega$. Making a
corresponding substitution in Eq.~(\ref{eq:Gdotp}) we arrive at the
following equation for the non-equilibrium generalization of the
chiral vortical effect:
\begin{equation}
  \label{eq:cve}
  \bm j_\textrm{CVE}
 = \bm\omega \int_{\bm p}  2|\bm p| f (\bm{\hat p\cdot b})
=
 \frac{\bm \omega}{4\pi^2}\int_0^\infty
  \overline{f(E,\bm{\hat p})}\, 2E dE \,.
\end{equation}
This result is also in agreement with Ref.~\cite{Loganayagam:2012pz}
for isotropic distribution, and
reduces to $\bm j_\textrm{CVE}=\mu^2 \bm \omega/(2\pi)^2$ for the
well-known case of the Fermi-Dirac distribution at zero temperature.

\prlsec{Conclusion}
We presented kinetic description of the chiral magnetic
and chiral vortical effects given by kinetic equation~(\ref{eq:dfdtC})
with equations of motion~(\ref{eq:xdot}),~(\ref{eq:pdot}). Although
these equations are ubiquitous in the condensed matter literature on
the effects of the Berry curvature, to our knowledge, their
relationship to the chiral magnetic and chiral vortical effects has
not been appreciated until now. The key observation that the Berry
curvature for the Weyl Hamiltonian is sourced by a monopole at $|\bm
p|=0$ leads directly to the general non-equilibrium expressions for
the CME and CVE in Eqs.~(\ref{eq:jcme}) and~(\ref{eq:cve}) which
reproduce all known results in equilibrium.

The presence of the monopole
singularity in the momentum space also provides a natural mechanism by
which anomaly can change the particle number in an otherwise
classical system. The classical description breaks down in the region
surrounding the singularity at $\bm p=0$ of the size of order of the typical
momentum in the lowest Landau orbit. The net particle creation occurs by
the purely quantum effect of anomaly (level crossing) inside this
non-classical region. The net flux of the particles into the classical
region is then given by Eq.~(\ref{eq:S-Delta}), which can serve as a
boundary condition for the kinetic equation in the classical region.

It would be interesting to use the results obtained here to
investigate the consequences of non-equilibrium for the chiral
transport effects in heavy-ion collisions
\cite{Fukushima:2008xe,Fukushima:2010vw,Kharzeev:2010gr}. We leave
this and other applications to further study.

\acknowledgments
The authors thank R.~Loganayagam, P.~Surowka for discussions and
acknowledge the stimulating environment of the ``CPODD''
workshop at RIKEN-BNL Research Center. Y.Y. thanks
the James S.~Kouvel Fellowship Foundation for support.
This work is supported by
the DOE grant No.\ DE-FG0201ER41195.

\bibliography{kinetic}

\end{document}